\documentclass[11pt]{article}
\usepackage{amsmath}
\usepackage{amssymb}

\def\PsfigVersion{1.9}
\ifx\undefined\psfig\else\endinput\fi

\let\LaTeXAtSign=\@
\let\@=\relax
\edef\psfigRestoreAt{\catcode`\@=\number\catcode`@\relax}
\catcode`\@=11\relax
\newwrite\@unused
\def\ps@typeout#1{{\let\protect\string\immediate\write\@unused{#1}}}
\ps@typeout{psfig/tex \PsfigVersion}

\def\figurepath{./}

\def\@nnil{\@nil}
\def\@empty{}
\def\@psdonoop#1\@@#2#3{}
\def\@psdo#1:=#2\do#3{\edef\@psdotmp{#2}\ifx\@psdotmp\@empty \else
    \expandafter\@psdoloop#2,\@nil,\@nil\@@#1{#3}\fi}
\def\@psdoloop#1,#2,#3\@@#4#5{\def#4{#1}\ifx #4\@nnil \else
       #5\def#4{#2}\ifx #4\@nnil \else#5\@ipsdoloop #3\@@#4{#5}\fi\fi}
\def\@ipsdoloop#1,#2\@@#3#4{\def#3{#1}\ifx #3\@nnil 
       \let\@nextwhile=\@psdonoop \else
      #4\relax\let\@nextwhile=\@ipsdoloop\fi\@nextwhile#2\@@#3{#4}}
\def\@tpsdo#1:=#2\do#3{\xdef\@psdotmp{#2}\ifx\@psdotmp\@empty \else
    \@tpsdoloop#2\@nil\@nil\@@#1{#3}\fi}
\def\@tpsdoloop#1#2\@@#3#4{\def#3{#1}\ifx #3\@nnil 
       \let\@nextwhile=\@psdonoop \else
      #4\relax\let\@nextwhile=\@tpsdoloop\fi\@nextwhile#2\@@#3{#4}}
\ifx\undefined\fbox
\newdimen\fboxrule
\newdimen\fboxsep
\newdimen\ps@tempdima
\newbox\ps@tempboxa
\long\def\fbox#1{\leavevmode\setbox\ps@tempboxa\hbox{#1}\ps@tempdima\fboxrule
    \advance\ps@tempdima \fboxsep \advance\ps@tempdima \dp\ps@tempboxa
   \hbox{\lower \ps@tempdima\hbox
  {\vbox{\hrule height \fboxrule
          \hbox{\vrule width \fboxrule \hskip\fboxsep
          \vbox{\vskip\fboxsep \box\ps@tempboxa\vskip\fboxsep}\hskip 
                 \fboxsep\vrule width \fboxrule}
                 \hrule height \fboxrule}}}}
\fi
\newread\ps@stream
\newif\ifnot@eof       
\newif\if@noisy        
\newif\if@atend        
\newif\if@psfile       
{\catcode`\%=12\global\gdef\epsf@start{
\def\epsf@PS{PS}
\def\epsf@getbb#1{%
\openin\ps@stream=#1
\ifeof\ps@stream\ps@typeout{Error, File #1 not found}\else
   {\not@eoftrue \chardef\other=12
    \def\do##1{\catcode`##1=\other}\dospecials \catcode`\ =10
    \loop
       \if@psfile
          \read\ps@stream to \epsf@fileline
       \else{
          \obeyspaces
          \read\ps@stream to \epsf@tmp\global\let\epsf@fileline\epsf@tmp}
       \fi
       \ifeof\ps@stream\not@eoffalse\else
       \if@psfile\else
       \expandafter\epsf@test\epsf@fileline:. \\%
       \fi
          \expandafter\epsf@aux\epsf@fileline:. \\%
       \fi
   \ifnot@eof\repeat
   }\closein\ps@stream\fi}%
\long\def\epsf@test#1#2#3:#4\\{\def\epsf@testit{#1#2}
                        \ifx\epsf@testit\epsf@start\else
\ps@typeout{Warning! File does not start with `\epsf@start'.  It may not be a PostScript file.}
                        \fi
                        \@psfiletrue} 
{\catcode`\%=12\global\let\epsf@percent=
\long\def\epsf@aux#1#2:#3\\{\ifx#1\epsf@percent
   \def\epsf@testit{#2}\ifx\epsf@testit\epsf@bblit
        \@atendfalse
        \epsf@atend #3 . \\%
        \if@atend       
           \if@verbose{
                \ps@typeout{psfig: found `(atend)'; continuing search}
           }\fi
        \else
        \epsf@grab #3 . . . \\%
        \not@eoffalse
        \global\no@bbfalse
        \fi
   \fi\fi}%
\def\epsf@grab #1 #2 #3 #4 #5\\{%
   \global\def\epsf@llx{#1}\ifx\epsf@llx\empty
      \epsf@grab #2 #3 #4 #5 .\\\else
   \global\def\epsf@lly{#2}%
   \global\def\epsf@urx{#3}\global\def\epsf@ury{#4}\fi}%
\def\epsf@atendlit{(atend)} 
\def\epsf@atend #1 #2 #3\\{%
   \def\epsf@tmp{#1}\ifx\epsf@tmp\empty
      \epsf@atend #2 #3 .\\\else
   \ifx\epsf@tmp\epsf@atendlit\@atendtrue\fi\fi}

\chardef\psletter = 11 
\chardef\other = 12

\newif \ifdebug 
\newif\ifc@mpute 
\c@mputetrue 

\let\then = \relax
\def\r@dian{pt }
\let\r@dians = \r@dian
\let\dimensionless@nit = \r@dian
\let\dimensionless@nits = \dimensionless@nit
\def\internal@nit{sp }
\let\internal@nits = \internal@nit
\newif\ifstillc@nverging
\def \Mess@ge #1{\ifdebug \then \message {#1} \fi}

{ 
        \catcode `\@ = \psletter
        \gdef \nodimen {\expandafter \n@dimen \the \dimen}
        \gdef \term #1 #2 #3%
               {\edef \t@ {\the #1}
                \edef \t@@ {\expandafter \n@dimen \the #2\r@dian}%
                \t@rm {\t@} {\t@@} {#3}%
               }
        \gdef \t@rm #1 #2 #3%
               {{%
                \count 0 = 0
                \dimen 0 = 1 \dimensionless@nit
                \dimen 2 = #2\relax
                \Mess@ge {Calculating term #1 of \nodimen 2}%
                \loop
                \ifnum  \count 0 < #1
                \then   \advance \count 0 by 1
                        \Mess@ge {Iteration \the \count 0 \space}%
                        \Multiply \dimen 0 by {\dimen 2}%
                        \Mess@ge {After multiplication, term = \nodimen 0}%
                        \Divide \dimen 0 by {\count 0}%
                        \Mess@ge {After division, term = \nodimen 0}%
                \repeat
                \Mess@ge {Final value for term #1 of 
                                \nodimen 2 \space is \nodimen 0}%
                \xdef \Term {#3 = \nodimen 0 \r@dians}%
                \aftergroup \Term
               }}
        \catcode `\p = \other
        \catcode `\t = \other
        \gdef \n@dimen #1pt{#1} 
}

\def \Divide #1by #2{\divide #1 by #2} 

\def \Multiply #1by #2
       {{
        \count 0 = #1\relax
        \count 2 = #2\relax
        \count 4 = 65536
        \Mess@ge {Before scaling, count 0 = \the \count 0 \space and
                        count 2 = \the \count 2}%
        \ifnum  \count 0 > 32767 
        \then   \divide \count 0 by 4
                \divide \count 4 by 4
        \else   \ifnum  \count 0 < -32767
                \then   \divide \count 0 by 4
                        \divide \count 4 by 4
                \else
                \fi
        \fi
        \ifnum  \count 2 > 32767 
        \then   \divide \count 2 by 4
                \divide \count 4 by 4
        \else   \ifnum  \count 2 < -32767
                \then   \divide \count 2 by 4
                        \divide \count 4 by 4
                \else
                \fi
        \fi
        \multiply \count 0 by \count 2
        \divide \count 0 by \count 4
        \xdef \product {#1 = \the \count 0 \internal@nits}%
        \aftergroup \product
       }}

\def\r@duce{\ifdim\dimen0 > 90\r@dian \then   
                \multiply\dimen0 by -1
                \advance\dimen0 by 180\r@dian
                \r@duce
            \else \ifdim\dimen0 < -90\r@dian \then  
                \advance\dimen0 by 360\r@dian
                \r@duce
                \fi
            \fi}

\def\Sine#1%
       {{%
        \dimen 0 = #1 \r@dian
        \r@duce
        \ifdim\dimen0 = -90\r@dian \then
           \dimen4 = -1\r@dian
           \c@mputefalse
        \fi
        \ifdim\dimen0 = 90\r@dian \then
           \dimen4 = 1\r@dian
           \c@mputefalse
        \fi
        \ifdim\dimen0 = 0\r@dian \then
           \dimen4 = 0\r@dian
           \c@mputefalse
        \fi
        \ifc@mpute \then
                \divide\dimen0 by 180
                \dimen0=3.141592654\dimen0
                \dimen 2 = 3.1415926535897963\r@dian 
                \divide\dimen 2 by 2 
                \Mess@ge {Sin: calculating Sin of \nodimen 0}%
                \count 0 = 1 
                \dimen 2 = 1 \r@dian 
                \dimen 4 = 0 \r@dian 
                \loop
                        \ifnum  \dimen 2 = 0 
                        \then   \stillc@nvergingfalse 
                        \else   \stillc@nvergingtrue
                        \fi
                        \ifstillc@nverging 
                        \then   \term {\count 0} {\dimen 0} {\dimen 2}%
                                \advance \count 0 by 2
                                \count 2 = \count 0
                                \divide \count 2 by 2
                                \ifodd  \count 2 
                                \then   \advance \dimen 4 by \dimen 2
                                \else   \advance \dimen 4 by -\dimen 2
                                \fi
                \repeat
        \fi             
                        \xdef \sine {\nodimen 4}%
       }}

\def\Cosine#1{\ifx\sine\UnDefined\edef\Savesine{\relax}\else
                             \edef\Savesine{\sine}\fi
        {\dimen0=#1\r@dian\advance\dimen0 by 90\r@dian
         \Sine{\nodimen 0}
         \xdef\cosine{\sine}
         \xdef\sine{\Savesine}}}              

\def\psdraft{
        \def\@psdraft{0}
}
\def\psfull{
        \def\@psdraft{100}
}

\psfull

\newif\if@scalefirst
\def\psscalefirst{\@scalefirsttrue}
\def\psrotatefirst{\@scalefirstfalse}
\psrotatefirst

\newif\if@draftbox
\def\psnodraftbox{
        \@draftboxfalse
}
\def\psdraftbox{
        \@draftboxtrue
}
\@draftboxtrue

\newif\if@prologfile
\newif\if@postlogfile
\def\pssilent{
        \@noisyfalse
}
\def\psnoisy{
        \@noisytrue
}
\psnoisy
\newif\if@bbllx
\newif\if@bblly
\newif\if@bburx
\newif\if@bbury
\newif\if@height
\newif\if@width
\newif\if@rheight
\newif\if@rwidth
\newif\if@angle
\newif\if@clip
\newif\if@verbose
\def\@p@@sclip#1{\@cliptrue}

\newif\if@decmpr

\def\@p@@sfigure#1{\def\@p@sfile{null}\def\@p@sbbfile{null}
                \openin1=#1.bb
                \ifeof1\closein1
                        \openin1=\figurepath#1.bb
                        \ifeof1\closein1
                                \openin1=#1
                                \ifeof1\closein1%
                                       \openin1=\figurepath#1
                                        \ifeof1
                                           \ps@typeout{Error, File #1 not found}
                                                \if@bbllx\if@bblly
                                                \if@bburx\if@bbury
                                                        \def\@p@sfile{#1}%
                                                        \def\@p@sbbfile{#1}%
                                                        \@decmprfalse
                                                \fi\fi\fi\fi
                                        \else\closein1
                                                \def\@p@sfile{\figurepath#1}%
                                                \def\@p@sbbfile{\figurepath#1}%
                                                \@decmprfalse
                                        \fi%
                                \else\closein1%
                                        \def\@p@sfile{#1}
                                        \def\@p@sbbfile{#1}
                                        \@decmprfalse
                                \fi
                        \else
                                \def\@p@sfile{\figurepath#1}
                                \def\@p@sbbfile{\figurepath#1.bb}
                                \@decmprtrue
                        \fi
                \else
                        \def\@p@sfile{#1}
                        \def\@p@sbbfile{#1.bb}
                        \@decmprtrue
                \fi}

\def\@p@@sfile#1{\@p@@sfigure{#1}}

\def\@p@@sbbllx#1{
                \@bbllxtrue
                \dimen100=#1
                \edef\@p@sbbllx{\number\dimen100}
}
\def\@p@@sbblly#1{
                \@bbllytrue
                \dimen100=#1
                \edef\@p@sbblly{\number\dimen100}
}
\def\@p@@sbburx#1{
                \@bburxtrue
                \dimen100=#1
                \edef\@p@sbburx{\number\dimen100}
}
\def\@p@@sbbury#1{
                \@bburytrue
                \dimen100=#1
                \edef\@p@sbbury{\number\dimen100}
}
\def\@p@@sheight#1{
                \@heighttrue
                \dimen100=#1
                \edef\@p@sheight{\number\dimen100}
}
\def\@p@@swidth#1{
                \@widthtrue
                \dimen100=#1
                \edef\@p@swidth{\number\dimen100}
}
\def\@p@@srheight#1{
                \@rheighttrue
                \dimen100=#1
                \edef\@p@srheight{\number\dimen100}
}
\def\@p@@srwidth#1{
                \@rwidthtrue
                \dimen100=#1
                \edef\@p@srwidth{\number\dimen100}
}
\def\@p@@sangle#1{
                \@angletrue
                \edef\@p@sangle{#1} 
}
\def\@p@@ssilent#1{ 
                \@verbosefalse
}
\def\@p@@sprolog#1{\@prologfiletrue\def\@prologfileval{#1}}
\def\@p@@spostlog#1{\@postlogfiletrue\def\@postlogfileval{#1}}
\def\@cs@name#1{\csname #1\endcsname}
\def\@setparms#1=#2,{\@cs@name{@p@@s#1}{#2}}
\def\ps@init@parms{
                \@bbllxfalse \@bbllyfalse
                \@bburxfalse \@bburyfalse
                \@heightfalse \@widthfalse
                \@rheightfalse \@rwidthfalse
                \def\@p@sbbllx{}\def\@p@sbblly{}
                \def\@p@sbburx{}\def\@p@sbbury{}
                \def\@p@sheight{}\def\@p@swidth{}
                \def\@p@srheight{}\def\@p@srwidth{}
                \def\@p@sangle{0}
                \def\@p@sfile{} \def\@p@sbbfile{}
                \def\@p@scost{10}
                \def\@sc{}
                \@prologfilefalse
                \@postlogfilefalse
                \@clipfalse
                \if@noisy
                        \@verbosetrue
                \else
                        \@verbosefalse
                \fi
}
\def\parse@ps@parms#1{
                \@psdo\@psfiga:=#1\do
                   {\expandafter\@setparms\@psfiga,}}
\newif\ifno@bb
\def\bb@missing{
        \if@verbose{
                \ps@typeout{psfig: searching \@p@sbbfile \space  for bounding box}
        }\fi
        \no@bbtrue
        \epsf@getbb{\@p@sbbfile}
        \ifno@bb \else \bb@cull\epsf@llx\epsf@lly\epsf@urx\epsf@ury\fi
}       
\def\bb@cull#1#2#3#4{
        \dimen100=#1 bp\edef\@p@sbbllx{\number\dimen100}
        \dimen100=#2 bp\edef\@p@sbblly{\number\dimen100}
        \dimen100=#3 bp\edef\@p@sbburx{\number\dimen100}
        \dimen100=#4 bp\edef\@p@sbbury{\number\dimen100}
        \no@bbfalse
}
\newdimen\p@intvaluex
\newdimen\p@intvaluey
\def\rotate@#1#2{{\dimen0=#1 sp\dimen1=#2 sp
                  \global\p@intvaluex=\cosine\dimen0
                  \dimen3=\sine\dimen1
                  \global\advance\p@intvaluex by -\dimen3
                  \global\p@intvaluey=\sine\dimen0
                  \dimen3=\cosine\dimen1
                  \global\advance\p@intvaluey by \dimen3
                  }}
\def\compute@bb{
                \no@bbfalse
                \if@bbllx \else \no@bbtrue \fi
                \if@bblly \else \no@bbtrue \fi
                \if@bburx \else \no@bbtrue \fi
                \if@bbury \else \no@bbtrue \fi
                \ifno@bb \bb@missing \fi
                \ifno@bb \ps@typeout{FATAL ERROR: no bb supplied or found}
                        \no-bb-error
                \fi
                \count203=\@p@sbburx
                \count204=\@p@sbbury
                \advance\count203 by -\@p@sbbllx
                \advance\count204 by -\@p@sbblly
                \edef\ps@bbw{\number\count203}
                \edef\ps@bbh{\number\count204}
                \if@angle 
                        \Sine{\@p@sangle}\Cosine{\@p@sangle}
                        {\dimen100=\maxdimen\xdef\r@p@sbbllx{\number\dimen100}
                                            \xdef\r@p@sbblly{\number\dimen100}
                                            \xdef\r@p@sbburx{-\number\dimen100}
                                            \xdef\r@p@sbbury{-\number\dimen100}}
                        \def\minmaxtest{
                           \ifnum\number\p@intvaluex<\r@p@sbbllx
                              \xdef\r@p@sbbllx{\number\p@intvaluex}\fi
                           \ifnum\number\p@intvaluex>\r@p@sbburx
                              \xdef\r@p@sbburx{\number\p@intvaluex}\fi
                           \ifnum\number\p@intvaluey<\r@p@sbblly
                              \xdef\r@p@sbblly{\number\p@intvaluey}\fi
                           \ifnum\number\p@intvaluey>\r@p@sbbury
                              \xdef\r@p@sbbury{\number\p@intvaluey}\fi
                           }
                        \rotate@{\@p@sbbllx}{\@p@sbblly}
                        \minmaxtest
                        \rotate@{\@p@sbbllx}{\@p@sbbury}
                        \minmaxtest
                        \rotate@{\@p@sbburx}{\@p@sbblly}
                        \minmaxtest
                        \rotate@{\@p@sbburx}{\@p@sbbury}
                        \minmaxtest
                        \edef\@p@sbbllx{\r@p@sbbllx}\edef\@p@sbblly{\r@p@sbblly}
                        \edef\@p@sbburx{\r@p@sbburx}\edef\@p@sbbury{\r@p@sbbury}
                \fi
                \count203=\@p@sbburx
                \count204=\@p@sbbury
                \advance\count203 by -\@p@sbbllx
                \advance\count204 by -\@p@sbblly
                \edef\@bbw{\number\count203}
                \edef\@bbh{\number\count204}
}
\def\in@hundreds#1#2#3{\count240=#2 \count241=#3
                     \count100=\count240        
                     \divide\count100 by \count241
                     \count101=\count100
                     \multiply\count101 by \count241
                     \advance\count240 by -\count101
                     \multiply\count240 by 10
                     \count101=\count240        
                     \divide\count101 by \count241
                     \count102=\count101
                     \multiply\count102 by \count241
                     \advance\count240 by -\count102
                     \multiply\count240 by 10
                     \count102=\count240        
                     \divide\count102 by \count241
                     \count200=#1\count205=0
                     \count201=\count200
                        \multiply\count201 by \count100
                        \advance\count205 by \count201
                     \count201=\count200
                        \divide\count201 by 10
                        \multiply\count201 by \count101
                        \advance\count205 by \count201
                     \count201=\count200
                        \divide\count201 by 100
                        \multiply\count201 by \count102
                        \advance\count205 by \count201
                     \edef\@result{\number\count205}
}
\def\compute@wfromh{
                \in@hundreds{\@p@sheight}{\@bbw}{\@bbh}
                \edef\@p@swidth{\@result}
}
\def\compute@hfromw{
                \in@hundreds{\@p@swidth}{\@bbh}{\@bbw}
                \edef\@p@sheight{\@result}
}
\def\compute@handw{
                \if@height 
                        \if@width
                        \else
                                \compute@wfromh
                        \fi
                \else 
                        \if@width
                                \compute@hfromw
                        \else
                                \edef\@p@sheight{\@bbh}
                                \edef\@p@swidth{\@bbw}
                        \fi
                \fi
}
\def\compute@resv{
                \if@rheight \else \edef\@p@srheight{\@p@sheight} \fi
                \if@rwidth \else \edef\@p@srwidth{\@p@swidth} \fi
}
\def\compute@sizes{
        \compute@bb
        \if@scalefirst\if@angle
        \if@width
           \in@hundreds{\@p@swidth}{\@bbw}{\ps@bbw}
           \edef\@p@swidth{\@result}
        \fi
        \if@height
           \in@hundreds{\@p@sheight}{\@bbh}{\ps@bbh}
           \edef\@p@sheight{\@result}
        \fi
        \fi\fi
        \compute@handw
        \compute@resv}

\def\psfig#1{\vbox {
        \ps@init@parms
        \parse@ps@parms{#1}
        \compute@sizes
        \ifnum\@p@scost<\@psdraft{
                \special{ps::[begin]    \@p@swidth \space \@p@sheight \space
                                \@p@sbbllx \space \@p@sbblly \space
                                \@p@sbburx \space \@p@sbbury \space
                                startTexFig \space }
                \if@angle
                        \special {ps:: \@p@sangle \space rotate \space} 
                \fi
                \if@clip{
                        \if@verbose{
                                \ps@typeout{(clip)}
                        }\fi
                        \special{ps:: doclip \space }
                }\fi
                \if@prologfile
                    \special{ps: plotfile \@prologfileval \space } \fi
                \if@decmpr{
                        \if@verbose{
                                \ps@typeout{psfig: including \@p@sfile.Z \space }
                        }\fi
                        \special{ps: plotfile "`zcat \@p@sfile.Z" \space }
                }\else{
                        \if@verbose{
                                \ps@typeout{psfig: including \@p@sfile \space }
                        }\fi
                        \special{ps: plotfile \@p@sfile \space }
                }\fi
                \if@postlogfile
                    \special{ps: plotfile \@postlogfileval \space } \fi
                \special{ps::[end] endTexFig \space }
                \vbox to \@p@srheight sp{
                        \hbox to \@p@srwidth sp{
                                \hss
                        }
                \vss
                }
        }\else{
                \if@draftbox{           
                        \hbox{\frame{\vbox to \@p@srheight sp{
                        \vss
                        \hbox to \@p@srwidth sp{ \hss \@p@sfile \hss }
                        \vss
                        }}}
                }\else{
                        \vbox to \@p@srheight sp{
                        \vss
                        \hbox to \@p@srwidth sp{\hss}
                        \vss
                        }
                }\fi

        }\fi
}}
\psfigRestoreAt
\let\@=\LaTeXAtSign

\makeatletter%
\def\nottoobig#1{{\hbox{$\left#1\vcenter to1.111\ht\strutbox{}\right.\n@space$}}}
\makeatother%

\topsep 8pt plus2pt minus4pt   

\makeatletter%
\def\@begintheorem#1#2{\trivlist\item[\hskip\labelsep{\bf #1\ #2}]}
\makeatother
\makeatletter 
\newcommand{\C}[1]{ {\rm {#1}} }
\newcommand{\band}{\bigwedge}
\newcommand{\Band}[3]{(\bigwedge#1\!\!:\,#2\!\!:\,#3)}
\newcommand{\bor}{\bigvee}
\newcommand{\Bor}[3]{(\bigvee#1\!\!:\,#2\!\!:\,#3)}
\newcommand{\Forall}[3]{(\forall #1\!\!:\,#2\!\!:\,#3)}
\newcommand{\Exists}[3]{(\exists #1\!\!:\,#2\!\!:\,#3)}
\newcommand{\Union}[3]{(\bigcup #1\!\!:\,#2\!\!:\,#3)}
\def\union{\,\bigcup\limits\,}
\newcommand{\true}{\mbox{\it true}}
\newcommand{\false}{\mbox{\it false}}
\newcommand{\SUM}[3]{ (\sum #1 \!\! : \, #2 \!\!:\, #3) }
\newcommand{\IFS}{\mbox{\bf if}}
\newcommand{\IF}[1]{ \mbox{\bf if} \, #1 \, \rightarrow \, }
\newcommand{\GC}[2]{ #1 \, \rightarrow \, #2 }
\newlength{\filength}
\settowidth{\filength}{\mbox{\bf f{}i}}
\newsavebox{\gcbox}
\sbox{\gcbox}{\framebox[\filength]{\rule{0ex}{2ex}}}
\newcommand{\BB}[1]{\usebox{\gcbox}\; #1 \, \rightarrow \, }
\newcommand{\FI}{\; \mbox{\bf f{}i}}
\newcommand{\Skip}{ \mbox{\bf skip} }
\newcommand{\DOS}{\mbox{\bf do}}
\newcommand{\DO}[1]{\mbox{\bf do} \, #1 \, \rightarrow \,}
\newcommand{\OD}{\mbox{\bf od}}
\newcommand{\cobegin}{{\bf cobegin}\,}
\renewcommand{\|}{\, //  \,}
\newcommand{\coend}{\,{\bf coend}}
\newcommand{\Set}[1]{ \hbox{\bf\{} #1 \hbox{\bf\}}}
\newcommand{\Bag}[1]{ \{\!| #1  |\!\}}
\newlength{\leftjustindent}
\newlength{\@leftjustindent}
\setlength{\@leftjustindent}{\leftmargin}
\def\leftjust{\let\\\@leftjustcr\let\end\@endleftjust
  \addtolength{\@leftjustindent}{\leftjustindent}
  \vcenter\bgroup
  \halign\bgroup
    \hbox to\displaywidth{
      \rule{\@leftjustindent}{0ex}$\displaystyle##$\hfill
      }\crcr
}
\def\endleftjust{\crcr\egroup\egroup\endgroup}
\def\@endleftjust#1{\crcr\egroup\egroup\@checkend{#1}\endgroup}
\def\@leftjustcr{\crcr}

\newcommand{\hoare}[3]{\{{#1}\}\:{#2}\:\{{#3}\}}
\renewcommand{\wp}[2]{ {\it wp}({#1},{#2})}
\newcommand{\assert}[1]{\!\{#1\}}
\newcommand{\atom}[1]{\langle\,{#1}\,\rangle}
\newcommand{\lbl}[1]{{#1 \!:\;\,}}
\newcommand{\pre}[1]{ {\it pre}({#1})}
\newcommand{\post}[1]{ {\it post}({#1}) }
\newcommand{\NI}[2]{ {\it NI}({#1},{#2}) }
\newcommand{\equi}[3]{ {  {\rm E}_{#1}^{#2}({#3}) }    }
\newcommand{\red}[3]{ {  {\rm R}_{#1}^{#2}({#3}) }    }
\newcommand{\sparse}{{{\rm SPARSE}}}
\newcommand{\tally}{{{\rm TALLY}}}
\newcommand{\inferfrom}[2]{\begin{array}[t]{c}\displaystyle
   \frac{#1}{#2}\end{array}}
\newtheorem{theorem}{Theorem}[section]

\newtheorem{corollary}[theorem]{Corollary}

\newcommand{\qedblob}{\mbox{\rule[-1.5pt]{5pt}{10.5pt}}}
\def\literalqed{{\ \nolinebreak\hfill\mbox{\qedblob\quad}}}
\def\qedcareful{\literalqed}
\def\qed{\literalqed}
\def\trueloveqed{{\ \nolinebreak\hfill\mbox{\boldmath
\Huge$ \Box$}\nolinebreak\mbox{$\!\!\!\!\!\!
{}^{\normalsize\heartsuit}$}}}

\newtheorem{fact}[theorem]{Fact}

\newcommand{\singlespacing}{\let\CS=
\@currsize\renewcommand{\baselinestretch}{1}\tiny\CS}
\newcommand{\singlespacingplus}{\let\CS=
\@currsize\renewcommand{\baselinestretch}{1.25}\tiny\CS}
\newcommand{\doublespacing}{\let\CS=
\@currsize\renewcommand{\baselinestretch}{1.75}\tiny\CS}
\newcommand{\draftspacing}{\let\CS=
\@currsize\renewcommand{\baselinestretch}{1.65}\tiny\CS}
\makeatother

\hyphenation{theory area areas theorem theorems par-allel par-allelize par-allelized threshold Hemaspaan-dra}

\newtheorem{definition}[theorem]{Definition}

\flushbottom{}
\makeatletter
\clubpenalty=\@highpenalty
\widowpenalty=\@highpenalty
\makeatother

\let\BLS=\baselinestretch
\emergencystretch=2em

\makeatletter
\newcommand{\niceonespacing}{\let\CS=\@currsize\renewcommand{\baselinestretch}{1.1}\tiny\CS}\newcommand{\nicetwospacing}{\let\CS=\@currsize\renewcommand{\baselinestretch}{1.2}\tiny\CS}
\newcommand{\nicethreespacing}{\let\CS=\@currsize\renewcommand{\baselinestretch}{1.3}\tiny\CS}
\newcommand{\singlespacingplusplus}{\let\CS=\@currsize\renewcommand{\baselinestretch}{1.35}\tiny\CS}
\newcommand{\nicefivespacing}{\let\CS=\@currsize\renewcommand{\baselinestretch}{1.5}\tiny\CS}
\newcommand{\nicesixspacing}{\let\CS=\@currsize\renewcommand{\baselinestretch}{1.6}\tiny\CS}
\newcommand{\nicefoospacing}{\let\CS=\@currsize\renewcommand{\baselinestretch}{1.7}\tiny\CS}
\newcommand{\normalspacing}{\doublespacing}
\makeatother

\makeatletter
\def\@cite#1#2{[#1\if@tempswa , #2\fi]}
\makeatother

\makeatletter
\def\@citex[#1]#2{\if@filesw\immediate\write\@auxout{\string\citation{#2}}\fi
  \def\@citea{}\@cite{\@for\@citeb:=#2\do
    {\@citea\def\@citea{,\linebreak[0]}\@ifundefined
       {b@\@citeb}{{\bf ?}\@warning
       {Citation `\@citeb' on page \thepage \space undefined}}%
\hbox{\csname b@\@citeb\endcsname}}}{#1}}
\makeatother

\newcommand{\proofendsign}{$\rule{2mm}{2mm}$}
\newenvironment{proof}{{\noindent \bf Proof }}{{\hspace*{\fill}\proofendsign\par\bigskip}}
\newcommand{\eqq}[1]{\lq\lq #1\rq\rq}
\newcommand{\eq}[1]{\lq #1\rq}
\newcommand{\todopic}{{\begin{picture}(5,5)\thicklines\put(0,0){\line(1,0){5}}\put(0,0){\line(1,2){2.5}}\put(5,0){\line(-1,2){2.5}}\put(2.5,2){\makebox(0,0){{\rm !}}}\end{picture}}}
\newcommand{\todo}[1]{{\setlength{\unitlength}{1mm}\todopic}\footnote{{\setlength{\unitlength}{0.8mm}\todopic} #1}}
\newcommand{\card}[1]{||#1||}
\newcommand{\redstyle}[1]{\mathnormal{#1}}
\newcommand{\reduction}[2]{\redstyle{\le_{\mathrm{#2}}^{\mathrm{#1}}}}
\newcommand{\polyreduction}[1]{\,\reduction{p}{#1}\,}

\newcommand{\redm}{\polyreduction{m}}
\newcommand{\redctt}{\polyreduction{c}}
\newcommand{\reddtt}{\polyreduction{d}}
\newcommand{\redbtt}{\polyreduction{btt}}
\newcommand{\redbttctt}{\polyreduction{btt(c)}}
\newcommand{\redcttbtt}{\polyreduction{c(btt)}}
\newcommand{\reddttbtt}{\polyreduction{d(btt)}}
\newcommand{\redtt}{\polyreduction{tt}}
\newcommand{\redpa}{\polyreduction{\parallel}}
\newcommand{\redT}{\polyreduction{T}}
\newcommand{\redkahorn}{\polyreduction{k{\scriptscriptstyle \!-\!}ah}}

\newcommand{\sharpp}{{\rm \#P}}
\newcommand{\sharpsat}{{\rm \#SAT}}
\newcommand{\sat}{{\rm SAT}}
\newcommand{\usatq}{{\rm USAT_Q}}
\newcommand{\qbf}{{\rm QBF}}
\newcommand{\parityp}{{\rm \oplus P}}
\newcommand{\up}{{\rm UP}}
\newcommand{\us}{{\rm US}}
\newcommand{\fewnp}{{\rm FewNP}}
\newcommand{\fewp}{{\rm FewP}}
\newcommand{\coup}{{\rm coUP}}
\newcommand{\e}{{\rm E}}
\renewcommand{\exp}{{\rm EXP}}
\newcommand{\NE}{{\rm NE}}
\renewcommand{\ne}{{\rm NE}}
\newcommand{\nexp}{{\rm NEXP}}
\newcommand{\p}{{\rm P}}
\newcommand{\littlep}{{\rm p}}
\newcommand{\NP}{{\rm NP}}
\newcommand{\FP}{{\rm FP}}
\newcommand{\npnp}{{\rm NP^{NP}}}
\newcommand{\bh}{{\rm BH}}
\newcommand{\BH}{{\rm BH}}
\newcommand{\BPP}{{\rm BPP}}
\newcommand{\Prob}{{\rm Prob}}
\newcommand{\MOD}{{\rm MOD}}
\newcommand{\BPTIME}{{\rm BPTIME}}
\newcommand{\ZPTIME}{{\rm ZPTIME}}
\newcommand{\DTIME}{{\rm DTIME}}
\newcommand{\dtime}{{\rm DTIME}}
\newcommand{\BPSPACE}{{\rm BPSPACE}}
\newcommand{\charfunc}[1]{\chi_{#1}}
\newcommand{\der}{\vdash} \newcommand{\notder}{\not\vdash}
\newcommand{\packalgo}{\mathrm{PackAlg}}
\newcommand{\listalgo}{\mathrm{ListAlg}}
\newcommand{\rhs}[1]{\mathrm{RHS}(#1)}
\newcommand{\ie}{{\mbox{i.e.}}}
\newcommand{\inter}{{\cap}}
\newcommand{\spp}{{\rm SPP}}
\newcommand{\gapp}{{\rm GapP}}
\newcommand{\pl}{{\rm PL}}
\def\sstar{\Sigma^{*}}
\newcommand{\R}{{\rm R}}

\newcommand{\np}{{\rm NP}}
\newcommand{\nt}{{\rm NT}}
\newcommand{\nnt}{{\rm NNT}}
\newcommand{\parityoptp}{{\rm \oplus{}OptP}}
\newcommand{\optp}{{\rm OptP}}
\newcommand{\diffp}{{\rm D^P}}
\newcommand{\pp}{{\rm PP}}
\newcommand{\bpp}{{\rm BPP}}
\newcommand{\zpp}{{\rm ZPP}}
\newcommand{\cor}{{\rm coR}}
\newcommand{\npc}{$\np$-com\-plete}
\newcommand{\conp}{{\rm coNP}}
\newcommand{\pspace}{{\rm PSPACE}}
\newcommand{\eespace}{{\rm EESPACE}}
\newcommand{\dspace}{{\rm DSPACE}}
\newcommand{\psp}{{\pspace}}
\newcommand{\pnexp}{{\p^\nexp}}
\newcommand{\npnexp}{{\np^\nexp}}
\newcommand{\nenp}{{\ne^\np}}
\newcommand{\enp}{{\e^\np}}
\newcommand{\pnp}{{\p^\np}}
\newcommand{\pnplog}{{\p^{\np[\log ]}}}
\newcommand{\pij}{{\p^{\bh_i:\bh_j}}}
\newcommand{\pji}{{\p^{\bh_j:\bh_i}}}
\newcommand{\nexpnp}{{\nexp^\np}}
\newcommand{\coNP}{{\rm coNP}}
\newcommand{\cone}{{\rm CONE}}
\newcommand{\sigmatwozero}{{\Sigma_2^0}}
\newcommand{\pitwozero}{{\Pi_2^0}}
\newcommand{\pithreezero}{{\Pi_3^0}}
\newcommand{\sigmathreezero}{{\Sigma_3^0}}
\newcommand{\sigmatwo}{{\Sigma_2^{\littlep}}}
\newcommand{\sigmathree}{{\Sigma_3^{\littlep}}}
\newcommand{\sigmafour}{{\Sigma_4^{\littlep}}}
\newcommand{\sigmafive}{{\Sigma_5^{\littlep}}}
\newcommand{\sigmak}{{\Sigma_k^{\littlep}}}
\newcommand{\sigmai}{{\Sigma_i^{\littlep}}}
\newcommand{\sigmaj}{{\Sigma_j^{\littlep}}}
\newcommand{\pitwo}{{\Pi_2^{\littlep}}}
\newcommand{\pithree}{{\Pi_3^{\littlep}}}
\newcommand{\pifour}{{\Pi_4^{\littlep}}}
\newcommand{\pifive}{{\Pi_5^{\littlep}}}
\newcommand{\thetatwo}{{\Theta_2^{\littlep}}}
\newcommand{\deltatwo}{{\Delta_2^{\littlep}}}
\newcommand{\poly}{{\rm poly}}
\newcommand{\ph}{{\rm PH}}
\newcommand{\few}{{\rm Few}}
\newcommand{\fewch}{{\rm FewCH}}
\newcommand{\eh}{{\rm EH}}
\def\bull{\vrule height .9ex width .8ex depth -.1ex }
\newcommand{\blob}{\mbox{\rule[-1.5pt]{5pt}{10.5pt}}}
\newcommand{\lindent}{\qquad}
\newcommand{\magicnum}{{ n^{\frac{1-\epsilon}{\epsilon}+\delta}}}
\newcommand{\fsup}{{\,f_{super}\,}}
\newcommand{\fred}{{\,f_{reduced}\,}}
\newcommand{\pne}{{\p^\ne}}
\newcommand{\npne}{{\np^\ne}}
\newcommand{\nnexarg}{{\nxx^\nexx (x) }}
\newcommand{\nnexx}{{\nxx^\nexx  }}
\newcommand{\nnex}{{\nxx^\nexx }}
\newcommand{\expnp}{{\exp^\np }}
\newcommand{\nxx}{{\rm N_{17}}}
\newcommand{\nexx}{{\rm NE_{21}}}
\newcommand{\seh}{{\rm SEH}}
\newcommand{\sexph}{{\rm SEXPH}}
\newcommand{\pstar}{{\p_\star}}
\newcommand{\nestar}{{\ne_{\,\star}}}
\newcommand{\supersetproper}{  \stackrel{\scriptscriptstyle\superset}{\scriptscriptstyle\not-}}
\newcommand{\subsetproper}{  \stackrel{\scriptscriptstyle\subset}{\scriptscriptstyle\not-}}
\newcommand{\superset}{\supset}
\newcommand{\superseteq}{\supseteq}

\newcommand{\substar}{\mbox{$\subset^*$}}
\newcommand{\superstar}{\mbox{$\superset^*$}}

\def\unionfromc{\,\textstyle\bigcup_{\scriptstyle c}\,}
\def\unionfromk{\,\textstyle\bigcup_{\scriptstyle k}\,}

\newcommand{\newlozenge}{\setlength{\fboxsep}{0pt}\setlength{\fboxrule}{.7pt}\framebox[6pt]{\rule{0pt}{9pt}}}

\def\pair#1{{{\langle\!\!~#1~\!\!\rangle}}}
\def\pairs#1{{{\langle\!\!~#1~\!\!\rangle}}}
\newcommand{\piso}{\mbox{$\littlep$-iso\-mor\-phic}}
\newcommand{\manyonea}{\mbox{$\,\leq_{\rm m}^{{\littlep},\,A}\,$}}
\newcommand{\manyone}{\mbox{$\,\leq_{\rm m}^{{\littlep}}$\,}}
\newcommand{\Turing}{\mbox{$\,\leq_{\rm T}^{{\littlep}}$\,}}
\newcommand{\paiso}{\mbox{$\littlep^A$-iso\-mor\-phic}}
\newcommand{\pisoa}{\paiso}
\newcommand{\pisoam}{\mbox{$\littlep^A$-iso\-mor\-phism}}
\newcommand{\pisom}{\mbox{$\littlep$-iso\-mor\-phism}}
\newcommand{\pselective}{\mbox{$\p$-selec\-tive}}
\newcommand{\sigmastar}{\mbox{$\Sigma^\ast$}}
\newcommand{\pisnp}{\mbox{$\p=\np$}}
\newcommand{\usuba}{\mbox{$U_A$}}
\newcommand{\univsuba}{\mbox{$Univ_A$}}
\newcommand{\pisnotnp}{\mbox{$\p\neq\np$}}
\newcommand{\lb}{\mbox{\{}}
\newcommand{\rb}{\mbox{\}}}
\newcommand{\pa}{\mbox{$\p^A$}}
\newcommand{\calf}{\mbox{$\cal F$}}
\newcommand{\calc}{\mbox{$\cal C$}}
\newcommand{\cald}{\mbox{$\cal D$}}
\newcommand{\calcone}{{\cal C}_1}
\newcommand{\calctwo}{{\cal C}_2}
\newcommand{\npa}{\mbox{$\np^A$}}
\newcommand{\conpa}{\mbox{$\conp^A$}}
\newcommand{\upa}{\mbox{$\up^A$}}
\newcommand{\sparses}{\mbox{ sparse $S\,$}}
\newcommand{\bigo}{\mbox{$\cal O$}}
\newcommand{\condition}{\,\nottoobig{|}\:}
\def\land{{\; \wedge \;}}

\newcommand{\parallelnp}{\mbox{$\p_{||}^{\np}$}}
\newcommand{\rp}{\rm R}
\newcommand{\corp}{{\rm coR}}
\newcommand{\ceqp}{{\rm C_{\!=\!}P }}
\newcommand{\pclose}{\rm P-close}
\newcommand{\apt}{\rm APT}
\newcommand{\ppoly}{\rm P/poly}
\newcommand{\dr}{\mbox{\tt Carroll Ranking}}
\newcommand{\dw}{\mbox{\tt Carroll Winner}}
\newcommand{\ds}{\mbox{\tt Carroll Score}}
\newcommand{\mee}{\mbox{\tt MEE}}
\sloppy

\def\nats{\naturalnumber}
\newcommand{\naturalnumber}{\ensuremath{{  \mathbb{N} }}}
\def\wit#1{{\mbox{\rm{}WIT}_M(#1)}}
\newcommand{\acomp}{\mbox{\em acomp}}

\begin{document}

\title{A Moment of Perfect Clarity I:\\
The Parallel Census Technique\footnote{\protect\singlespacing
Supported in part 
by grant
NSF-INT-9815095/\protect\linebreak[0]DAAD-315-PPP-g{\"u}-ab
and the Studienstiftung des Deutschen Volkes.
Written in part while the second author was
visiting 
Julius-Maximilians-Universit\"at.}}

\author{Christian Gla{\ss}er\footnote{\protect\singlespacing
E-mail: {\tt glasser@informatik.uni-wuerzburg.de}.}
\\
  Theoretische Informatik \\
Institut f\"ur Informatik \\
  Julius-Maximilians-Universit\"at\\
 Am Hubland \\
 97074 W\"{u}rzburg \\
 Germany
\and
Lane A. Hemaspaandra\footnote{\protect\singlespacing
E-mail: {\tt lane@cs.rochester.edu}.}
\\
Department of Computer Science \\
University of Rochester \\
Rochester, NY 14627 \\
USA}

\date{July 13, 2000}

{\singlespacing

\singlespacing\maketitle

}

\begin{abstract}
    We discuss the history and uses of the
    parallel census technique---an elegant tool in the study of
    certain computational objects having polynomially bounded census
    functions. 
    A sequel~\cite{gla-hem:SPECIALjtoappearANDinprep:clarityII} 
    will discuss advances (including \cite{cai-nai-siv:tCarefulMostlyOutdatedBySTACSbutNpDttSparseImpliesUSATQinPisImplicitOnlyHereSeeComments:sparse-hard}
    and Gla{\ss}er \cite{gla:t:sparse}), some related to the parallel 
    census technique and some due to other approaches, 
    in the complexity-class
    collapses that follow if $\NP$ has sparse hard sets under
    reductions weaker than (full) truth-table reductions.
\end{abstract}

\section{Introduction}
\label{ALT:sec_parallel_census_technique}

Have you ever used a pair of binoculars? Then you know the process one
goes through to initially set the distance between the 
two eyepieces---sometimes the view may black out, yet 
if one goes too far the other way 
one has two circles of view that don't coincide. However, there is a
point where things are just right: All is crisply aligned and one can
enjoy the view of that pileated woodpecker, at
least if it has been so polite as to
wait while one was playing with the interocular adjustment.

The parallel census technique is very much like this: too far one way
and our view blacks out, too far the other way and we get chaos from
overlapping views, but at a certain magic point everything comes into
focus.

\section{The Parallel Census Technique}
So, what {\em is\/} 
the parallel census technique? Loosely put---and since we
are speaking of a flavor of approach loosely put is appropriate here---the 
parallel census technique refers to an approach that
can be used when faced with a deterministically or
nondeterministically recognizable type of objects of which one knows
that one has at most polynomially many but one does not know exactly how
many one has. The parallel census approach is to, in parallel, for
each possible guess of the cardinality to ask, also in parallel, a
bunch of questions to a nondeterministic set (machine) that will guess
and check a number of objects matching the cardinality you guessed and
that will test some bit of information about one of the objects. Our
set of 
questions will be such that the questions corresponding to the
correct cardinality will---when all their answers are viewed
together---reveal everything about the objects: the overall
cardinality and even the name of each object.

Put more succinctly, the parallel census technique is a way of gaining
information via parallelized queries to nondeterministic classes. For
example the following known result can be crisply seen via the
parallel census technique.

\begin{definition}
A polynomial-time truth-table reduction 
(see~\cite{lad-lyn-sel:j:com} 
or~\cite{gla-hem:SPECIALjtoappearANDinprep:clarityII})
is said to be {\em exponentially length-decreasing\/}
if there is a constant $k$ such that, on all inputs $x$,
each of the (at most polynomial number of) parallel queries
generated on input $x$ is of length 
at most $k \log |x|$.
\end{definition}

\begin{theorem} \label{thm_sparse_reduces_ne}
    (can be seen via the tools of \cite{har:j:sparse,har-imm-sew:j:sparse},
    see the discussion in Section~\ref{s:attribution})~~Every
    sparse $\NP$ set reduces via a parallel (i.e., truth-table), 
    exponentially length-decreasing
    reduction to an $\NE$ set 
    ($\NE = \bigcup_{k} \mathrm{NTIME}[2^{kn}]$).
\end{theorem}

\begin{definition}
    \cite{val-vaz:j:np-unique}
    Let $Q(\cdot)$ be a one-argument boolean predicate. Let $\sat$ be,
    as usual, the set of satisfiable boolean formulas.
    $$\usatq(x) = { \left\{
        \begin{array}{lll}
            \sat(x) & ~& \mbox{if the number of satisfying assignments of $x$ is $0$ or $1$} \\  & & \\
            Q(x) & & \mbox{otherwise.}
        \end{array}
    \right.}$$
\end{definition}

\begin{definition} \label{def_sharpp_fewp_few}~
    \begin{itemize}
        \item[{a)}] {\cite{val:j:enumeration}} A
            function $f$ is in $\sharpp$ iff there is some NPTM, $N$, such
            that for each $x$ it holds that $N(x)$
            (i.e., the computation of $N$ on input $x$)
            has exactly $f(x)$
            accepting paths.
        \item[{b)}] {\cite{val:j:checking}} A set $B$ is in
            $\up$ iff there is a $\sharpp$
            function $f$ satisfying
            $(\forall x) [f(x) \le 1]$ such that $B = \{ x \condition
            f(x) > 0 \}$.
        \item[{c)}] {\cite{all-rub:j:print}} A set $B$ is in
            $\fewp$ iff there is a $\sharpp$
            function $f$ satisfying $(\exists \mbox{ polynomial }q)
            (\forall x) [f(x) \le q(|x|)]$
            such that $B = \{ x \condition
            f(x) > 0 \}$.
        \item[{d)}] {\cite{cai-hem:j:parity}}
            A set $B$ is in $\few$ iff there is a polynomial-time
            predicate $R(\cdot,\cdot)$ and a $\sharpp$ function $f$
            satisfying $(\exists \mbox{ polynomial }q) (\forall x) [f(x)
            \le q(|x|)]$ such that $B = \{ x \condition R(x,f(x)) \}$.
        \end{itemize}
\end{definition}

\begin{fact} \label{fact_p_up_fewp_few}
    $\p \subseteq \up \subseteq \fewp \subseteq \few$.
\end{fact}

\begin{theorem} \label{thm_few_reduces_usatq}
    (see Section~\ref{s:attribution} for a discussion of
    attribution)~~If $L \in \few$ then $(\forall Q) [L \redtt \usatq]$,
    where $\redtt$~\cite{lad-lyn-sel:j:com}
    denotes truth-table (i.e., parallel) reductions.
\end{theorem}

We defer a discussion of the history and attribution of
Theorems~\ref{thm_sparse_reduces_ne} and \ref{thm_few_reduces_usatq}
until after the proof of Theorem~\ref{thm_few_reduces_usatq}. The
proof of Theorem~\ref{thm_few_reduces_usatq} will provide a
quintessential example of the parallel census technique.

\begin{proof}{\bf (Theorem~\ref{thm_few_reduces_usatq}):}~~Let 
$L$ be an arbitrary set from $\few$. Let $R(\cdot, \cdot)$,
    $f$, and $q(\cdot)$ be as in Definition~\ref{def_sharpp_fewp_few}(d).
    With respect to $\sharpp$ function $f$, let $N$ be as in 
Definition~\ref{def_sharpp_fewp_few}(a), 
     and without loss of
    generality assume that for each $x$ all computation paths of
    $N(x)$ are of length exactly $p(|x|)$, where $p$ is a polynomial
    and $(\forall w) [p(w) \ge 1]$.

As we discuss at the end of 
the proof, the proof will actually
establish even the claim that 
the complete list of accepting paths of
$N(x)$ is computable in $\FP_{\mathrm{tt}}^{\usatq}$.

    Consider the set
    \begin{eqnarray*}
        J &=& \Big\{ \langle x,c,j,k,b \rangle \condition
            c \le q(|x|) \,\wedge\, 1 \le k \le p(|x|) \,\wedge\,
            1 \le j \le c \,\wedge\, b \in \{ 0,1 \} \,\wedge\, \\
        && \phantom{\{ \langle x,c,j,k,b \rangle \condition}
            (\exists ~ p_1 < p_2 < \cdots < p_c) \Big[
            (\mathrm{the\ } k^{\mathrm{th}} \mathrm{\ bit\ of\ }
            p_j \mathrm{\ is\ } b) \,\wedge\, \\
        && \phantom{\{ \langle x,c,j,k,b \rangle \condition}
            (\forall i : 1 \le i \le c) [p_i
            \mathrm{\ is\ an\ accepting\ path\ of\ } N(x)] \Big] \Big\}.
    \end{eqnarray*}
    Let $N'$ be the obvious, natural, $\np$ machine
    for $J$. Note that this machine will have the property that on
    each input of the form $\langle x, f(x), \cdot, \cdot, \cdot
    \rangle$, machine $N'$ will have either zero or one accepting
    path, as there will be just one valid guess for $(p_1, p_2,
    \ldots, p_c)$. Also, crucially, note that when $c$ is $f(x)$ then
    as $j$, $k$, and $b$ vary they basically read off all the bits of
    the accepting paths of $N(x)$. (Actually they poke each bit
    twice. This does no harm, and helps us avoid confusing \eqq{the
    accepting path having all zeros} with the lack of any accepting
    paths.)

    Let $\sigma_{N'}$ be a polynomial-time, {\em parsimonious}
    many-one reduction (Cook's reduction can be implemented
    parsimoniously \cite{gal:j:encodings}) from questions about
    membership in $J$ to boolean formulas. That is, for each string
    $v$, the formula $\sigma_{N'}(v)$ will have exactly as many
    satisfying assignments as $N'(v)$ had accepting computation paths.

    Let $Q$ be any predicate. The reduction to implement $L \redtt
    \usatq$ is as follows. On input $x$, ask $\usatq$, in parallel,
    all questions of the form $$\sigma_{N'}(x,c,j,k,b)$$ for all $c \le
    q(|x|)$, $1 \le j \le c$, $1 \le k \le p(|x|)$, $b \in \{0,1\}$.

    Note that the answers here are a bit magical. In particular,
    consider all answers associated (on input $x$) with a single value
    of $c$, say, $c'$. If $c'>f(x)$, that is, if $c'$ is greater than
    the number of accepting paths of $N(x)$, then all the questions
    \mbox{\eqq{$\sigma_{N'}(\langle x,c',j,k,b \rangle) 
    \in \usatq$?}}
    will
    get the answer \eqq{no} (since in this case 
   $N'(\langle x,c',j,k,b \rangle)$ will have
    zero accepting paths, $\sigma_{N'}(\langle x,c',j,k,b \rangle)$ 
    will be unsatisfiable,
    so $\usatq$ cannot contain it). Of course, if $c' < f(x)$, that
    is, if $c'$ is less than the number of accepting paths of $N(x)$,
    then the answers to the questions 
    \mbox{\eqq{$\sigma_{N'}(\langle
    x,c',j,k,b \rangle) \in \usatq$?}} will yield a big muddle of
    information, since the too-low guessed cardinality $c'$ will
    allow multiple valid guesses of $(p_1, p_2, \ldots, p_c)$ and so
    bits will be overlayed in ways that may potentially hide
    information. However, and this is the beautiful core of the
    parallel census technique, when $c' = f(x)$, that is, when $c'$
    equals the number of accepting paths of $N(x)$,
    then if $f(x) = 0$ every question of the
    form 
    \mbox{\eqq{$\sigma_{N'}(\langle x,c',j,k,b \rangle) \in \usatq$?}}
    will get the answer \eqq{no,} and if $f(x) > 0$ then at least one
    question of the form 
    \mbox{\eqq{$\sigma_{N'}(\langle x,c',j,k,b \rangle)
    \in \usatq$?}} will get the answer \eqq{yes,} and furthermore the
    answer to all questions of the form \mbox{\eqq{$\sigma_{N'}(\langle
    x,c',j,k,b \rangle) \in \usatq$?}} will in effect specify the
    entire list of accepting paths! That is, when $c' = f(x)$
    everything comes into focus---a moment of perfect clarity.

    So, after we get back the huge list of answers in parallel, if all
    the answers are \eqq{no} we know that $N(x)$ has zero accepting
    paths, and we accept exactly if $R(x,0)$. If at least one answer
    is yes, we consider the largest $\hat{c}$ for which some query
    \eqq{$\sigma_{N'}(\langle x,\hat{c},j,k,b \rangle) \in \usatq$?}
    received the answer \eqq{yes.} So, $\hat{c} = f(x)$. So, we accept
    exactly if $R(x,\hat{c})$. This already proves our stated theorem,
    but note that a bit more holds. From the whole collection of
    answers to the questions \eqq{$\sigma_{N'}(\langle x,c',j,k,b
    \rangle) \in \usatq$?} we can reconstruct all the accepting paths
    of $N(x)$. So not only is it true that $L \redtt \usatq$, but in
    fact it even holds that the complete list of accepting paths of
    $N(x)$ is computable in $\FP_{\mathrm{tt}}^{\usatq}$.
\end{proof}

From the very general claim expressed as 
Theorem~\ref{thm_few_reduces_usatq} one
can immediately conclude many of the 
other ways that the benefits of the
parallel census technique are used as they relate to $\few$. Most
particularly, it is immediate from Theorem~\ref{thm_few_reduces_usatq}
and Fact~\ref{fact_p_up_fewp_few} that the following corollary holds.

\begin{corollary} \label{coro_usatq_fewp}
    If $(\exists Q)[\usatq \in \p]$ then $\p = \few$ (and thus $\p =
    \up$ and $\p = \fewp$).
\end{corollary}

\section{Comments and Attributions}\label{s:attribution}

We come to the issue of
attribution. This is a bit tricky, as where one sees this as
originating depends in part on how flexible one is in defining
\eqq{this.} However, in terms of pointing 
to where the parallel census technique came to be seen as a key
approach to reaching conclusions about $\fewp$, that is relatively
clear: Selman (\cite{sel:t:adaptive} and its journal version
\cite{sel:j:taxonomy}; see especially Proposition~6 of the former,
which is Theorem~2 of the latter) and Toda (the proof of
Theorem~3.10 of \cite{tod:j:psel}).

The result $(\exists Q) [\usatq \in \p] \Rightarrow \p = \fewp$ is
explicitly stated by Buhrman, Fortnow, and Torenvliet
\cite{buh-for-tor:c:SixHypothesis}, and is there attributed as
\eqq{this was essentially proven in Toda's paper \cite{tod:j:psel}.}
Since it is now known that $\p = \fewp \Longleftrightarrow \p = \few$
\cite{sel:j:taxonomy}, in light of this equivalence one can conclude
Corollary~\ref{coro_usatq_fewp} from
this. Theorem~\ref{thm_few_reduces_usatq} may never have been stated
before in the strong form it appears here, but it just reflects an attempt
to distill to its core what is going on in the parallel census
technique of Selman and Toda.

We mention that the fundamental machinery needed to exploit the idea
behind the parallel census technique appears in a 1983 paper of
Hartmanis (Theorem~2.1 of \cite{har:j:sparse}, and it reappears in
Hartmanis, Immerman, and Sewelson \cite{har-imm-sew:j:sparse} as
Theorem~1; see 
also~\cite{sew:thesis:np}).\footnote{\protect\singlespacing
Upward separation (equivalently referred to sometimes
as ``downward collapse''), the  theme of the work of
Hartmanis~\protect\cite{har:j:sparse}
and Hartmanis, Immerman, and Sewelson~\protect\cite{har-imm-sew:j:sparse},
has been further studied---qualified, extended, etc.---in 
such papers as
\protect\cite{har-yes:j:computation,all-wil:j:downward,all:j:lim,rao-rot-wat:j:upward,hem-jha:j:defying,bei-gol:j:beta,hem-hem-hem:j:downward-translation,buh-for:j:two-queries,hem-hem-hem:j:easy-hard-survey}.}
We say that the \eqq{fundamental machinery} is there as in
both \cite{har:j:sparse} and \cite{har-imm-sew:j:sparse} the proofs
build sequential algorithms. However, if one closely examines the
proofs, it becomes clear that (since the number of census values is
polynomial) one can redo the algorithm so that it works via one big
round of parallel queries. If one were to do this, what one would
arrive at would be what is stated earlier in this paper as
Theorem~\ref{thm_sparse_reduces_ne}. 

The name \eqq{parallel census technique} was coined in Arvind et 
al.~\cite{arv-han-hem-koe-loz-mun-ogi-sch-sil-thi:b:sparse}, which noted
the relation of the technique to the work of Hartmanis, Immerman, and
Sewelson. Arvind et 
al.~\cite{arv-han-hem-koe-loz-mun-ogi-sch-sil-thi:b:sparse} also somewhat
improves on the number of queries from Selman~\cite{sel:t:adaptive}, and
generalizes the technique beyond $\np$.

Finally, as it will be useful in 
Part~II's~\cite{gla-hem:SPECIALjtoappearANDinprep:clarityII} discussion of
sparse sets hard for $\np$ with respect to various reductions, we
state the following famous result of Valiant and Vazirani
\cite{val-vaz:j:np-unique}. This result gives a quite different
conclusion from the hypothesis $(\exists Q)[\usatq \in \p]$ than does
Corollary~\ref{coro_usatq_fewp}.

\begin{theorem} \label{usatq_r_np}
    \cite{val-vaz:j:np-unique}
    If $(\exists Q)[\usatq \in \p]$ then $\R = \np$.
\end{theorem}

\bigskip
{\bf Acknowledgments}
The authors thank 
Matthias~Galota,
Edith Hemaspaandra,
Heinz~Schmitz,
Klaus~W.~Wagner,
and Gerd~Wechsung
for helpful discussions and comments.

\newcommand{\etalchar}[1]{$^{#1}$}

\end{document}